Obtaining a contradiction between local realism and
quantum mechanics using only one correlation function

Ramon Lapiedra
Departament d'Astronomia i Astrofísica i Observatori Astronòmic,
Universitat de València, 46100 Burjassot, València, Spain

A. Pérez
Departament de Física Teòrica and IFIC, Universitat de València-
CSIC, 46100-Burjassot, València, Spain.

**Abstract**
    An ensemble consisting on systems of two entangled spin *1/2* particles, all of them in the same global quantum state, are considered. The two spins are measured, each of them, on a fixed direction, at two randomly selected measurement times. Realism, plus locality and freedom of choice referred to these chosen times, are assumed. Then, from the sole correlation function related to the two measurements, without considering any Bell inequalities, a contradiction between these assumptions and quantum mechanics is found.

**1.- Introduction**
    Let us consider many systems of two entangled spin ½ particles, *1* and *2*, always in the same global state $\psi$. Let us perform two space-like separated spin measurements, in the directions ***a*** and ***b*:** the first one on particle *1* and the second one on particle *2*. We assume local realism. Then, for the corresponding measurement outcomes, say *S=±1*, we can write the following functions:

$$A=A(\lambda,a), \quad B=B(\lambda,b), \qquad (1)$$

where $\lambda$ stands for the realistic hidden variable, *A* referring to particle *1* and *B* to particle *2*. Then, we will have for the correlation function *<a,b>* of the product of values *A* and *B* associated to these two measurement directions, the following expression

$$<a,b> = \int d\lambda \rho(\lambda) A(\lambda,a) B(\lambda,b), \qquad (2)$$

where $\rho(\lambda)$ is the normalized probability density of the $\lambda$ distribution.

As it is well known, Bell, in his seminal paper [1], proved that there is a contradiction between quantum mechanics (QM), on the one hand, and local realism Eq. (1) plus *freedom of choice*, on the other hand. This freedom of choice postulate means that the measurement directions $a$ and $b$ are assumed to be uncorrelated with the $\lambda$ hidden variable, which leads to the fact that $\rho(\lambda)$ in (2) does not depend on $a$ or $b$. Bell obtained such a contradiction by first proving his celebrated inequalities for two entangled spin ½ particles in the singlet state, under the local realism and freedom of choice postulates, and then showing that these inequalities are violated by QM in the case of this particular state.

In the present paper we are going to show the existence of a contradiction between QM, on the one hand, and the local realism plus freedom of choice postulates, on the other hand. Here, locality and freedom refer not to the measuring directions, as usually, but to the measuring times (see next for details). To reach this contradiction, we will not need to consider any Bell inequalities, as we will only need to consider one correlation function such as (2), instead of the three different ones, associated to three different pairs of measurement directions, which are present in the seminal Bell inequalities.

## 2.- The contradiction

We still deal with the above pairs of spin measurements in the directions $a$ and $b$, respectively, where the two entangled spin ½ particles are always in the same initial state $\psi$, the number of these pairs being large enough.

But, differently to what was explained in the precedent section, we now explicitly write down the two random times $t_i$ and $t_j$, at which the two measurements of each pair are performed. These times are counted, for each pair, from the preparation time and can randomly repeat themselves. We then assume realism, not necessarily local in relation with the measuring directions, although we assume locality for the measuring times: That is, the $A$ measurement outcomes depend not only on its own measurement direction $a$, but might also depend on the other particle measurement direction $b$; similarly for the $B$ outcomes. However, we allow $A$ to depend on $t_i$ but not on $t_j$, and likewise for $B$. In all, we have to substitute Eqs. (1) by the following functions:

$$A = A(\lambda,a;b,t_i), \qquad B = B(\lambda,b;a,t_j). \tag{3}$$

It is to be remarked that, for our purpose in the present paper, we do not actually need this dependence on the measurement time since, as we will see next, we only need the possibility to label the different performed measurements by their corresponding measuring times. In any case, we will require that both measurement are space-like separated events, so that no effect can propagate from one event to the second one. This assumption makes it particularly plausible for the type of time dependence required in (3), i.e., that the $A$ function depends on $t_i$ but not on $t_j$ and likewise for the $B$ function. However, even with this space-like separation, we allowed in (3) that $A$ might depend on $\boldsymbol{b}$, and $B$ on $\boldsymbol{a}$, that is, in the parlance of [2], we would allow for "superdeterminism" or "supercorrelation".

For each measured pair we are assuming that, since its preparation and before the first measurement, the initial state evolves freely, and similarly for the corresponding collapsed state after this first measurement and before the second one.

Notice that, despite the inclusion of the new time arguments or labels, $t_i$, $t_j$, we still need to include the hidden variable value $\lambda$ in order to have two well defined functions like (3), since in practice the arguments $(a,t_i;b)$ do not allow us to make a true singular prediction of the corresponding measuring outcome, and the same for $(b,t_j;a)$.

Suppose that there are $N^2$ different randomly selected time pairs $(t_i,t_j)$, with $N$ large enough. Therefore, similarly to (2), we will have now for the corresponding correlation function $<a,b>$

$$<a,b> = (1/N^2)\sum_{(i,j)} \int d\lambda \rho_{(i,j)}(\lambda)A(\lambda,a;b,t_i)B(\lambda,b;a,t_j), \tag{4}$$

where the summation runs over the $N^2$ different $(i,j)$ pairs.

We now postulate freedom of choice for the different measuring time pairs, that is, we assume that $\rho_{(i,j)}(\lambda)$ does not depend on $(i,j)$, which means that we can substitute $\rho_{(i,j)}(\lambda)$ by $\rho(\lambda)$ in (4). Then, having in mind that in QM we have $<a,b> = -\boldsymbol{a.b}$, Eq. (4) becomes

$$\boldsymbol{a.b} = - (1/N^2) \sum_{(i,j)} \int d\lambda \rho(\lambda)A(\lambda,a;b,t_i)B(\lambda,b;a,t_j). \tag{5}$$

Equivalently,

$$\boldsymbol{a}.\boldsymbol{b} = - \int d\lambda \rho(\lambda)(1/N)\sum_i A(\lambda,a;b,t_i)(1/N)\sum_j B(\lambda,b;a,t_j), \qquad (6)$$

with the indices i and j running independently from $1$ to $N$. Let us take a fixed, albeit arbitrary $\lambda$ value, and calculate the corresponding partial sum $(1/N)\sum_i A(\lambda,a;b,t_i)$ in (6). If $N$ is large enough, we can identify this expression with the statistical average $<a>$, and write the following expression:

$$<a> \equiv (1/N)\sum_i A(\lambda,a;b,t_i) = \pi(\lambda,a^+) - \pi(\lambda,a^-), \qquad (7)$$

where $\pi(\lambda,a^+)$ is the probability of obtaining the value $A=+1$ in the $\boldsymbol{a}$ direction for this particular $\lambda$ value, whatever the corresponding $B$ outcome is, and similarly for $\pi(\lambda,a^-)$. Needless to say, in practice we are not able to select any given $\lambda$ value. Nevertheless, all $\lambda$ values share the same initial state $\psi$. This state is all we need in order to calculate the above probabilities as quantum probabilities. But these quantum probabilities are obviously independent of $\lambda$, only depending on $\psi$ and $\boldsymbol{a}$, and likewise for $<b> \equiv (1/N)\sum_j B(\lambda,b;a,t_j)$. Then, since $\int d\lambda \rho(\lambda) = 1$, we obtain from (6):

$$\boldsymbol{a}.\boldsymbol{b} = - <a><b>. \qquad (8)$$

an equality that cannot be satisfied in general, since the scalar product $\boldsymbol{a}.\boldsymbol{b}$ cannot always be reproduced by the uncorrelated product $-<a><b>$ . For instance, in the particular case when the initial state $\psi$ reduces to the singlet state, it can be easily seen that the right hand side of (8) vanishes, which is incompatible with the left hand side, unless $\boldsymbol{a}$ and $\boldsymbol{b}$ become orthogonal.

Thus QM, on the one hand, and realism plus freedom of choice and locality, both (freedom and locality) referred to the measuring times, on the other hand, cannot both be true, as we wanted to prove. Notice that, as previously announced, in attaining this result we have not used any Bell-like inequalities, since we only have dealt with just one correlation function $<a,b>$ instead of using, for instance, the tree different ones appearing in the seminal Bell inequalities.

The above contradiction proof can be made slightly more general by relaxing the kind of time dependence in (6) and writing instead a new expression

$$\boldsymbol{a}.\boldsymbol{b} = - (1/N^2) \sum_{(i,j)} \int d\lambda \rho(\lambda) A(\lambda,a;b,t_i) B(\lambda,b;a,t_j,t_i), \quad (9)$$

where now the $B$ function is allowed to depend on time $t_i$ too, by assuming $t_j > t_i$ (for any times $t_j$, $t_i$, belonging or not to the same measurement pair), and also the "induction" postulate [3,4] that makes $A$ independent of the later time $t_j$. As mentioned above in relation to Eq. (3), we demand that both measurements are space-like separated events. Then (9) becomes

$$\boldsymbol{a}.\boldsymbol{b} = - (1/N) \sum_i \int d\lambda \rho(\lambda) A(\lambda,a;b,t_i) \; (1/N)\sum_j B(\lambda,b;a,t_j,t_i), \quad (10)$$

while we have again, for the singlet initial state,

$$<b> \equiv (1/N)\sum_i B(\lambda,a;b,t_j,t_i) = 0, \quad (11)$$

this leading to the vanishing of the right hand side of (10). However, in general, this is not compatible with the left hand side value, $\boldsymbol{a}.\boldsymbol{b}$. In all, we are led to a new contradiction, this time between (9) and quantum mechanics, Eq. (9) being a consequence of both, the freedom of choice postulate (that makes $\rho(\lambda)$ independent of times $t_i$, $t_j$) and the above "induction" postulate (that makes function $A$ independent of $t_j$).

### 3.- Discussion

Thus, neither the locality, nor the freedom of choice loopholes, referred to the measuring directions, play any role in the reasoning of the present case. The first loophole is no more a necessary postulate to conclude the proof of the above contradiction with quantum mechanics, and similarly for the second one since in our argument we deal with only one correlation function relative to a unique pair of measurement directions $\boldsymbol{a}$ and $\boldsymbol{b}$, so that the problem of possible correlations among different $\lambda$ values and different pairs of measurement directions is no present any more: there are simply no <u>different</u> pairs of measurement directions in the present case.

Both the locality and the freedom of choice postulates, referred to the measuring times, can be made more plausible by assuming space-like separation between both pairs of measurements (for the plausibility of the locality postulate), and space–like separation between the selection of the two measurement times and the quantum preparations of the initial state $\psi$ (for the plausibility of freedom of choice postulate). As it is very well known, referred to the measurement directions, this first space-like separation was experimentally attained for the first time by Aspect [5]. The second kind of space-like separation, also referred to measurement directions, has been experimentally attained more recently by Scheidl *et al.* [6]. In both cases, the experiments involved photon polarization measurements instead of spin ½ measurements.

Notice that the assumed locality and freedom of choice, applied to the measurement times are more plausible assumptions than the similar ones frequently applied to the measurement directions: after all, the value of a correlation function such as (4) depends on the measurement directions, but not on the measurement times. This means that the functions $A$ and $B$ necessarily depend on the directions, but not necessarily on the times. Then, if we assume that $A$ and $B$ do not depend on the times, these times would become, as remarked in the precedent section, mere labels, which would straightly lead to the validity of both assumptions, locality and freedom of choice, applied to the measurement times.

In reference [7], in accordance with some general considerations presented in [8], it is claimed that, without resorting to the corresponding Bell inequalities, a correlation function such as the one in (2) leads, with the aid of quantum mechanics, to a detailed expression different to the one directly implied by quantum mechanics. However, the author does not carry his observation to its final conclusion, that is, the fact that the computed values of these two expressions are actually in mutual contradiction. He limits himself to claiming that the difference in the two expressions of the correlation function is the reason why Bell inequalities are violated by quantum mechanics. In the present paper we have been able to show that this contradiction between quantum mechanics, on the one hand, and realism plus locality and freedom of choice, the two last postulates applied to the measurement times, on the other hand, can already be seen at the level of just one correlation function, without any reference to Bell inequalities.

## Acknowledgements

This work was supported by the Spanish "Ministerio de Economía y Competividad", MICINN-FEDER project FIS-2012-33582 (R. Lapiedra), and by project FPA2011-23897 and "Generalitat Valenciana" grant GVPROMETEOII2014-087 (A. Pérez).